\documentclass[aps,prl,showpacs,showkeys,twocolumn]{revtex4-1}
\usepackage{graphicx,hyperref,amsmath,amsfonts}
\usepackage{epstopdf,color,bm,multirow,rotating,soul}

\begin{document}
\title{Spectral-Topological Superefficient Quantum Memory}
\author{N.S. Perminov$^{1}$, S.A. Moiseev$^{1, *}$}
\affiliation{$^{1}$Kazan Quantum Center, Kazan National Research Technical University n.a. A.N.Tupolev-KAI, 10 K. Marx, Kazan 420111, Russia}
\email{samoi@yandex.ru}

\begin{abstract}						
In this work we report about the fundamental correspondence between the high efficiency of the broadband quantum memory on multifrequency absorbers in the single-mode cavity and the topology of the observed frequency lines. 
We have found the spectral-topological matching condition by using the multiparametric optimization of the absorber characteristics which makes it possible to improve considerably the spectral efficiency of quantum memory  under conditions of a real experiment.
The elaborated approach opens the possibility for the creation of the broadband quantum interface consisting of a small number of the absorbers and characterized by the super-high quantum efficiency of more than 99.9\%.
\end{abstract}

\date{\today}

\maketitle

\textit{Introduction.}
The development of the optical quantum memory (QM) is of decisive importance for quantum information technologies \cite{Kurizki2015,Lvovsky2009,Hammerer2010,Kupriyanov2017,Ma2017}.
Impressive experimental results on the way to create the effective QM were achieved in the last decade \cite{Hedges2010,Hosseini2011,Cho2016,Chen_2013}.
At the same time, the further improvement of the quantum efficiency to values extremely close to 100\% that is necessary for the usage in quantum communications and especially in a quantum computer remains a complicated unsolved experimental problem.
In addition to a number of related tasks, first of all the solution of this problem requires the creation of the highly effective interface for the reversible storage of photons in long-lived coherent systems.

One of promising technological approaches is based on the realization of multimode QM on the reversible photon echo  \cite{Moiseev2001,Tittel2009} and on the idea of it implementation in the broadband optical or microwave cavities \cite{Afzelius2010,Moiseev2010,Sabooni2013_1,Sabooni2013_2,Jobez2014,Krimer2016,Arcangeli_2016}.
Owing to the enhancement of the interaction between the resonance system of atoms and light it was possible to increase considerably the efficiency and decrease the working number of atoms \cite{Sabooni2013_1,Sabooni2013_2,Jobez2014}.
The increase in the efficiency of the interface in this manner is also possible but only by providing the matching condition for a wide range of working frequencies \cite{Moiseev2010,Moiseev2013_1,EMoiseev2016}.
The general solution of this problem remains unknown in the present time that strongly hampers the search for practical ways of creating the highly effective broadband optical QM.

In this work, starting from the known AFC protocol of the photon echo QM \cite{Riedmatten2008} in the single mode cavity \cite{Afzelius2010}, we show that a system of a small number of resonant absorbers (emitters, atoms, particles, mini-resonators etc.) placed in a common cavity makes it possible to implement the super-high spectral efficiency larger than 99.9\% in the working frequency band.
We found that such high efficiency is implemented in the vicinity of parameters, where the topological restructuring of the spectrum of the system and the change of the number of observed resonance lines is recorded.
In this case the maximally possible efficiency is achieved due to the optimization of all parameters of resonant particles and the usage of the spectral-topological matching condition.

\textit{Physical model.}
The general theoretical model of the considered memory corresponds to the so-called impedance matching QM on the photon echo in a single mode cavity \cite{Moiseev2010,Afzelius2010,Moiseev2013_1,Kalachev2013}, which was expanded further on the system of ring resonators connected with the nanooptical fiber \cite{ESM-LPL-2017} and on other schemes \cite{EMoiseev2016,Yuan2016,Moiseev_2017_PRA}.
Using the input-output formalism of quantum optics \cite{Walls} for the single-photon field, for the system studied in the work we obtain equations for the excited modes of particles $s_n(t)$ and of the common cavity field $a(t)$:
\begin{align}\label{gen_eq_0} 
& \left[\partial_{t}+i\Delta_n+\gamma_n\right]s_n(t)+g_n^{0*}a(t)=0,\\
& \nonumber \left[\partial_{t}+\frac{\kappa}{2}\right]a(t)-\sum_ng_n^0s_n(t)=\sqrt{\kappa}a_{in}(t),
\end{align}
where $a_{in}(t)=[2\pi]^{-1/2}\int~d\nu e^{-i\nu t}f_\nu$ is the input pulse, $f_\nu$ is the spectral profile of the input pulse, for which the normalization condition for the single-photon field is fulfilled $\int d\nu |f_\nu|^2=1$, $\nu$ is the frequency counted from the central frequency of the radiation $\omega_0$ , $\Delta_n$ are frequency detuning of particles, $n\in\{-N,...,N\}\setminus\{0\}$, $\gamma_n$ is the attenuation decrement of the $n$-th particle, $\kappa$ is the coupling coefficient of the external waveguide with the broadband resonator, $g_n^0$ is the coupling constant of modes of the common resonator and the $n$-th particle.
We have also ignored Langevin forces \cite{Scully1997} in equations (\ref{gen_eq_0}) by focusing only to the searching of highest quantum efficiency in the studied scheme. 

From equations (\ref{gen_eq_0}) we obtain the output field $a_{out}(t)=\sqrt{\kappa}a(t)-a_{in}(t)$ in terms of the transfer function (TF) $S(\nu)=\tilde{a}_{out}(\nu)/\tilde{a}_{in}(\nu)$ in the form \begin{align}\label{gen_sol}
 S(\nu)=\frac{1+iF(\nu)}{1-iF(\nu)},
\end{align}
where
$F(\nu)=2\nu/\kappa+\sum_n g_n/(\Delta_n-i\gamma_n-\nu)$,
$a_{in,out}(t)=[2\pi]^{-1/2}\int~d\nu e^{-i\nu t} \tilde{a}_{in,out}(\nu)$, $2|g_n^0|^2/\kappa=g_n$ is the effective line width of a separate particle inside the common broadband resonator (for $\gamma_n=0$).
In the general case TF (\ref{gen_sol}) has a very complicated spectral behavior due to the strong interaction of the particles in the common cavity.
However, we show below that under certain conditions TF can acquire spectral properties corresponding to the nearly ideal QM.

\textit{Spectral-topological matching conditions.}
Introducing the delay time $T(\nu)=-i\operatorname{Arg}(S(\nu))/\nu$ on frequency $\nu$, we write TF in the form $S(\nu)=e^{i\nu T(\nu)}$.
In accordance with the general properties of the QM on a periodic frequency comb \cite{Riedmatten2008}, it is possible to formulate the main principle of obtaining the effective broadband QM: the broadband memory can be effective then and only then, when the delay time is the same for all frequencies in the given spectral range $\Omega$, i.e.,
\begin{align}\label{gen_top_cond}
T(\nu)=T(\nu_0) ~\forall ~ \nu\in \Omega,
\end{align}
where $\nu_0$ is the central frequency of the given range $\Omega$ (lower than $\nu_0=0$). 

It was found earlier that the condition (\ref{gen_top_cond}) can be fulfilled with the accuracy to terms $\sim \nu^4$ \cite{Moiseev2010,Moiseev2013_1} and $\sim \nu^6$ \cite{EMoiseev2016} in the vicinity $\nu=0$ that limits anyway the spectral range of the high quantum efficiency. 
Below we show that it is possible to reach the high quantum efficiency in a wider spectral range by reaching the higher accuracy of the fulfillment of the equality (\ref{gen_top_cond}). 
Imposing the larger number of conditions on physical parameters of the system (a set $\{g_n,\Delta_n\}$), we consider (\ref{gen_top_cond}) as equality in series:
\begin{align}\label{gen_top_cond_1}
\left\{\Bigl|\partial_\nu^\alpha(T(\nu)-T(\nu_0))\Bigr|_{\nu=\nu_0}\rightarrow\min\right\},
\end{align}
where $\alpha\in\{0,...,4N-1\}$ is determined by the maximal number of free parameters of the problem ($\{g_n,\Delta_n\}$).

In analytical calculations below we consider the case of small own losses of resonance particles $\gamma_n\ll1$ under the assumption of the fulfillment of the regime $"$broadband cavity$"$, when $N\Delta/\kappa\leq \sqrt{\gamma_n/\Delta}\ll 1$. With allowance for the symmetry  of the distribution of spectroscopic parameters $g_{-n}=g_n,~\Delta_{-n}=-\Delta_n$, which is necessary for the creation of the high quantum efficiency in the broad frequency band, from (\ref{gen_top_cond_1}), we  find the following algebraic system of $4N$ spectral-topological  matching conditions  on parameters $\{g_n,~\Delta_n\}$:
\begin{align}\label{phys_cond}
& \left|\sum\limits_{n=1}^N\frac{g_n}{\Delta_n^{2m+2}}-\frac{(2^{2m+2}-1)|B_{2m+2}|}{(2m+2)!}\left[T(0)\right]^{2m+1}\right|\rightarrow \min,\\
& \nonumber T(0)=4\sum\limits_{n=1}^N\frac{g_n}{\Delta_n^{2}},
\end{align}
where $m\in\{1,...,4N-1\}$, $B_m$ are Bernoulli numbers ($B_0=1,~B_2=1/6,...$). 

In fact, conditions (\ref{phys_cond}) are the statement of the problem of the multiparametric optimal control of spectral properties of the QM written in the algebraic form that makes it possible to apply algebraic geometry \cite{dolotin2007,Wyman1989,Martin1978,Previato2001} to search for the ways to improve the QM.
Below we show that the conditions of the implementation of the highly effective broadband QM (\ref{gen_top_cond}) are associated with the change of the topology of its spectrum and are fulfilled near the point of the topological spectral transition.

\textit{Topological transitions in the QM spectrum.}
For the case of the $2N$-particle system, when the initial frequency modes are detuned equidistantly $\Delta_n=\Delta(n-1/2)$ 
(further $\Delta=1$, i.e., the consideration is performed in units of $\Delta$), and the line widths of modes are the same $g_n=g$, 
we find $F(\nu)=2g\nu\sum\limits_{n=1}^{N}[(n-1/2)^2-\nu^2]^{-1}$. 
Further from the matching condition (\ref{phys_cond}) we obtain the following relationships for the optimal quantity $g=g_{cr}$ and the time of the signal recovery $T(0)$:
\begin{align}\label{wo_correct_cond}
& g_{cr}=\frac{\Delta}{\pi}\left[1-\frac{\psi^{(3)}\left(N-\frac{1}{2}\right)}{\pi^4}\right]^{\frac{1}{2}}\times\left[1-\frac{2\psi^{(1)}\left(N-\frac{1}{2}\right)}{\pi^2}\right]^{-\frac{3}{2}}\\
& \nonumber T(0)=\frac{2\pi}{\Delta}\times\frac{\pi g_{cr}}{\Delta}\times\left[1-\frac{2\psi^{(1)}\left(N-\frac{1}{2}\right)}{\pi^2}\right],
\end{align}
where $\psi^{(m)}(x)$ is the polygamma function. 
Expanding (\ref{wo_correct_cond}) over $\frac{1}{N}$ we have $\frac{\pi g_{cr}}{\Delta}\simeq 1+\frac{3}{\pi^2N}$ and
$T(0)\simeq\frac{2\pi^2g_{cr}}{\Delta^2}\left(1-\frac{2}{\pi^2N}\right)$. 
We see that for the optimal QM (at $g=g_{cr}$),
the difference in the echo time-of-flight $T(0)=\frac{2\pi}{\Delta}(1+\frac{1}{\pi^2N})$ is very small in comparison with the continuum case $T(0)=\frac{2\pi}{\Delta}$ ($N=\infty$) \cite{Riedmatten2008,Afzelius2010} for $N\gtrsim10$ but at the lower number of particles $N<10$ this difference becomes essential.

\begin{figure}[h]
\includegraphics[width=0.4\textwidth]{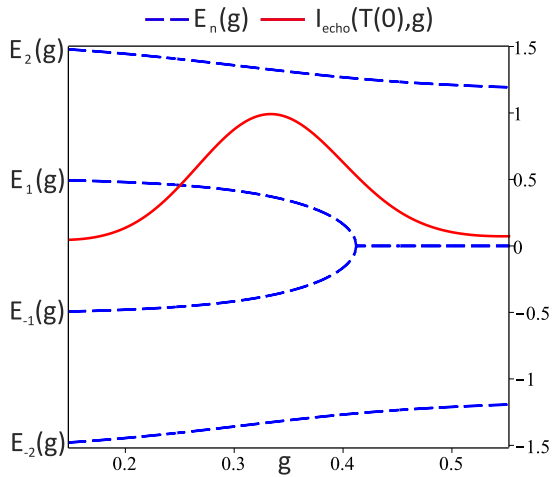}
\caption{Position of lines $E_{n}(g)$ of the TF spectrum of the four-particle system (blue dotted lines) and the normalized intensity $I_{echo}(T(0),g)$ of the recovered light pulse (red solid line) as a function of the coupling constant $g$.}
\label{E_n_Intens}
\end{figure}

The numerical analysis of the own modes TF (\ref{E_n_Intens}) carried out for the mode of the $"$broadband cavity$"$  ($\Delta/\kappa\ll1$) and weak relaxation ($\gamma/\Delta\ll 1$) shows that near the point of the maximal quantum efficiency ($g=g_{cr}$) the topological restructuring of the spectrum of the TF system occurs (see Fig. \ref{E_n_Intens}).
That is, at the rather weak coupling constant $g<g_{cr}$ the spectrum of TF resonance lines consists of $2N$ lines, and at the coupling strength $g\ge g_{cr}$ the number of lines is decreased by unity, i.e., it is $2N-1$.

The correspondence between the quantum efficiency and the change of the TF spectrum $E_{n}(g)$ is seen most clearly in the parametric space at the comparison of plots for the normalized intensity of the echo signal $I_{echo}(T(0),g)=|\int d\nu e^{-i\nu T(0)}S(\nu)f_{\nu}|^2/|\int d\nu f_{\nu}|^2$ ( $f_{\nu}=[2\pi\sigma^2]^{-1/4}e^{-\nu^2/(4\sigma^2)}$ -- the spectrum of the input pulse is chosen in the form of the Gaussian profile with the width $\sigma=0.2N\Delta$ and the central part of the spectrum of lines $E_{n}(g),~n\in\{-2,-1,1,2\}$. 
Fig. \ref{E_n_Intens} shows that the efficiency close to 100\% corresponds to the maximum of the echo signal intensity $I_{echo}(T(0),g)\rightarrow1$, which takes place in the region of restructuring of lines of the spectrum $E_{n}(g)$ in the small vicinity of the point of merging of two lines. 
The numerical calculations show that the pattern similar to Fig. \ref{E_n_Intens}, takes place for any even number of resonance particles.

The spectral-topological restructuring in the parametric space occurs in the rather small region of the variation of parameters and is the universal condition of the implementation of the high quantum efficiency irrespective of the certain form of the signal radiation.
The finer spectral optimization of the efficiency can depend on the used frequency band and the form of signal pulses, which, however, requires the additional study taking into account certain parameters of light fields analogous in meaning to that used in the scheme of the QM on slow light \cite{Gorshkov2007}.

\textit{Optimization of the efficiency in the wide frequency band.}
To study the properties of the QM in the wide spectral interval irrespective of the form of the signal we  introduce the function of TF spectral errors $\delta S^2(\nu)=|S(\nu)^2-S_0(\nu)^2|$, showing the deviation of TF from TF of the ideal memory $S_0(\nu)=e^{i\nu T(0)}$. 
For N=2 when the initial frequencies of particles are detuned equidistantly $\Delta_{\pm n}=\Delta(\pm n\mp1/2)$ (further $\Delta=1$, $n=1,2$), and the linewidths of modes are the same $g_n=g$, from (\ref{wo_correct_cond}) we obtain a set of spectroscopic data
$\{\Delta_{\pm 1} = \pm0.5,~\Delta_{\pm2} = \pm1.5,~g_{\pm1} = 0.318,~g_{\pm2} = 0.318\}$.
After the complete optimization according (\ref{phys_cond}) suppressing the negative spectral dispersion, we obtain the following topological structure of optimal parameters for frequency detuning and linewidths:
$\{\Delta_{\pm1} = \pm0.5,~\Delta_{\pm2} = \pm1.92,~g_{\pm1} = 0.318,~g_{\pm2} = 1.09\}$.

\begin{figure}[h]
\includegraphics[width=0.4\textwidth]{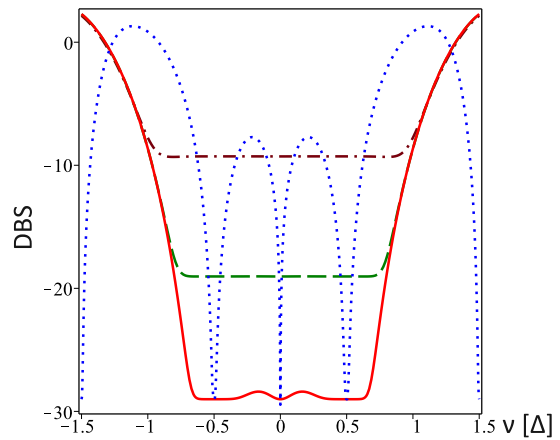}
\caption{Spectral error of TF in the log scale $DBS(\nu)=10\log_{10}(\delta S^2(\nu))$ for the four-particle system ($N=2$): red solid line –- the complete parameter optimization and $\gamma_n\sim 10^{-4}$, blue (dot) -- the partial optimization and $\gamma_n\sim 10^{-4}$, green (dash) -- the complete optimization and $\gamma_n\sim 10^{-3}$, brown (dot-dash) -- the complete optimization and $\gamma_n\sim 10^{-2}$.}
\label{4_res_dB_w_correction}
\end{figure}

It is seen from the results of the numerical calculation of Eq. (\ref{gen_sol}) given in Fig. \ref{4_res_dB_w_correction} that the comparison of the initial and optimized variants with allowance for own losses $\gamma_n\sim 10^{-4}$ achievable, e.g., upon using superconducting microwave resonators \cite{Brecht_2016} shows clearly the considerable improvement of spectral properties of the optimized variant ($\gamma_n\sim 10^{-4}$ corresponds to the quality factor $Q= 5\cdot 10^{6}$ of superconducting resonators for $\Delta=3\cdot 10^{7}$ and $\omega=3\cdot 10^{10}$).
Namely, in the second case the spectral quantum efficiency  $\eta(\nu)=|S(\nu)|^2$ weakly depends of the frequency in the spectral interval from $-0.6\Delta$ to $0.6\Delta$ and $\delta S^2(\nu)\sim 10^{-3}$ at $\gamma_n=10^{-4}$.
Thus, the optimization of parameters $\{g_n,\Delta_n\}$ makes it possible to create the almost ideal quantum interface  in this frequency region with the efficiency  $\eta(\nu)\cong 99.9\%$ as it seen in Fig. \ref{4_res_dB_w_correction} that is possible only upon using the controlled multifrequency system. 
Subsequent reversible transfer of the light field stored in the system of miniresonators to the long-lived electron--nuclear spin system \cite{Moiseev_2017_PRA} (for example in the rare--earth ions \cite{Zhong2015}) could provide on demand retrieval of the signal light, that is a subject of further studies.

The optimization of the topology of the spectrum of the highly effective QM can be also expanded to the wider spectral interval of frequencies by adding particles.
It will be necessary to correct parameters of all particles which is the reflection of their strong interaction.
For comparison, we note that the AFC protocol for the atomic ensemble in the optical resonator with four resonance lines was recently implemented experimentally \cite{Sabooni2013_1}, where the authors achieved the efficiency of 58\% record for the AFC protocol. 
The result obtained in the work \cite{Sabooni2013_1} can be dramatically improved in the approach we consider owing to the proposed optimization of frequency detuning and constants of the coupling of resonance lines.

\textit{Conclusion.}
The developed spectrally topological approach of the multiparametric optimization of the QM opens the possibility of creating broadband highly efficient multifrequency interface consisting of a small countable number of resonance particles (atoms, miniresonators etc.). 
That is, the spectral errors in the work of the interface can be decreased to extremely small values $\delta S^2(\nu)\sim 10^{-3}$ that meets the main technological requirements for the operation of the QM and its integration into quantum communication lines and the quantum  computer scheme.

An advantage of small-particle, energy-discrete resonance systems in comparison with spectrally continuous systems (atomic ensembles with the large inhomogeneous broadening) is in simpler possibilities of the perfect spectral control on the basis of using the existing optical and microwave technologies.
In optics the proposed approach can be implemented, e.g., in integral optical schemes containing systems of optical miniresonators connected with a nanooptical fiber \cite{Li2017}, where it is possible to select suitable frequencies of separate miniresonators and control the strength of their interaction with nanofibers \cite{O'Brien2009}, and the usage of several atoms with tunable frequencies being in the common resonator is also possible.
The usage of superconducting resonators connected with planar waveguides \cite{Brecht_2016} is the most technological in the microwave range frequency.

It is essential that the used algebraic approach \cite{dolotin2007} makes it possible to optimize all parameters of the considered system (\ref{phys_cond}) and gives an exhaustive answer to the question how it is possible to construct the superefficient quantum interface corresponding to the theoretical limit.

\textit{Acknowledgments.}
This work was financially supported by the Russian
Science Foundation through the Grant No. 14-12-01333-P.

\bibliographystyle{apsrev4-1}
\bibliography{ST_QM}

\end{document}